\documentclass{jpp}
\usepackage{graphicx}

\usepackage[utf8]{inputenc}
\usepackage[T1]{fontenc}
\usepackage{amsmath}
\usepackage{xcolor}

\shorttitle{Tightening energetic bounds}
\shortauthor{P. J. Costello and G. G. Plunk}

\author{P. J. Costello\aff{1}
  \corresp{\email{jpp@damtp.cam.ac.uk}},
  G. G. Plunk\aff{1}
 }
\affiliation{\aff{1}Max-Planck-Institut f\"ur Plasmaphysik, Wendelsteinstraße 1, 17491 Greifswald, Germany}

\title{Tightening energetic bounds on linear gyrokinetic instabilities}
\author{P. J. Costello \& G. G. Plunk}
\date{September 2024}

\begin{document}

\maketitle
\begin{abstract}
Bounding energetic growth of gyrokinetic instabilities is a complementary approach to linear instability analyses involving normal eigenmodes. Previous work has focused on upper bounds which are valid linearly and nonlinearly. However, if an upper bound on linear instability growth is desired, these nonlinearly valid bounds may be a poor predictor of the growth of the most unstable eigenmode. This is most evident for the simplest of instabilities: the ion-temperature-gradient (ITG) mode in slab geometry.
In this work, we derive energetic upper bounds specifically for linear instability growth, focusing on the slab ITG. We show that there is no fundamental limitation on how tightly linear growth can be bounded by an energetic norm, with the tightest possible bound being given by a special energy comprised of projection coefficients of the linear eigenmode basis. Additionally, we consider `constrained optimal modes' that maximise energy growth subject to constraints that are also obeyed by the linear eigenmodes. This yields computationally efficient upper bounds that closely resemble the linear growth rate, capturing effects connected to the real frequency of instabilities, which have been absent in the energetic bounds considered thus far.
\end{abstract}

\section{Introduction}

Most studies of gyrokinetic microinstabilities are concerned with \textit{linear eigenmodes} of the gyrokinetic equation, solutions that evolve in time as $\exp(- i \omega t)$, where $\omega = \omega_\mathrm{r} + i \gamma$ is a complex frequency. If $\gamma > 0$, these modes grow exponentially as $t \to \infty$ in the absence of any nonlinear interaction. The linear eigenmodes of the gyrokinetic equation are famously diverse and contribute to the richness of gyrokinetic theory but also complicate the computation of instability growth rates.  For instance it can be difficult to distinguish modes of different types \citep{Kammerer-merz-jenko-exceptional}, which react differently to changing plasma parameters. Recently, a new approach for studying microinstabilities has been developed in a series of papers \citep{helanderUpperBoundsGyrokinetic2021, helanderEnergeticBoundsGyrokinetic2022, plunkEnergeticBoundsGyrokinetic2023}, in which instability growth is instead bounded from above by the growth of \textit{optimal modes}. These modes are distinct from linear eigenmodes and are derived by choosing an energetic quadratic norm of the gyrokinetic system, say $E$, and seeking the distribution functions that maximise the normalised growth of this energy, $\Lambda = (2E)^{-1}\mathrm{d} E/ \mathrm{d} t$, instantaneously. The growth of the fastest growing optimal mode, $\Lambda_{\mathrm{max}}$, provides an upper bound on the allowable instantaneous growth of the system, and if the energetic norm considered is chosen carefully (such that it is invariant over the sum of nonlinear interactions) this bounds the growth of the linear system at each scale and bounds the nonlinear growth of the system when all scales are accounted for. The growth of the most unstable linear eigenmode at each scale necessarily satisfies,
 \begin{equation}
    \label{eqn:gamma_lambda_ineq}
     \gamma \leq \Lambda_{\mathrm{max}}.
 \end{equation} 
Although turbulence is a fundamentally nonlinear phenomenon, linear eigenmode analyses is often an informative endeavour, perhaps more than one would expect. This is evidenced by extensive gyrokinetic simulation work \citep{dannertGyrokineticSimulationCollisionless2005, pueschelStellaratorTurbulenceSubdominant2016, hatchSaturationGyrokineticTurbulence2011, hatchLinearSignaturesNonlinear2016} that shows that linear eigenmodes often survive nonlinearly. This observation forms the basis for quasilinear theory, which has proven to be a useful tool in modelling turbulent transport \citep{bourdelleNewGyrokineticQuasilinear2007}.

Until now, the gyrokinetic optimal modes have been computed with both linear and nonlinear growth in mind, often leading to large differences when the upper bound is compared to linear growth rates \citep{podaviniEnergeticBoundsGyrokinetic2024}. This is most evident for the classic case of instability driven by an ion-temperature gradient in a uniform magnetic field (i.e., the slab ITG mode), for which \cite{plunkEnergeticBoundsGyrokinetic2023} found that even a resonance-informed upper bound exhibited vastly different qualitative behaviour to the linear growth rate at small values of the temperature gradient. 
In this work, we explore ways to specialise the optimal mode analysis for the purpose of bounding linear instability growth, focusing specifically on the slab ITG mode, which exemplifies the shortcomings of the upper bounds considered thus far for the purpose of bounding linear instabilities.

Our work consists of two main parts. Firstly, we pose the question ``How tightly can linear eigenmode growth be bounded by optimal modes'' and find that the optimal mode growth spectrum can be made to correspond exactly to the growth rate spectrum of linear eigenmodes, giving a one-to-one correspondence between linear eigenmodes and optimal modes for this special energetic norm, which we refer to as the Case--Van Kampen energy.

With this theoretical result underfoot, we then seek simpler upper bounds on linear instability growth by considering what we refer to as \textit{constrained optimal modes}. This approach results in a low-dimensional, linear system of gyrofluid-like moment equations.  The growth of these constrained optimal modes exhibits a dependence on key instability parameters that is qualitatively very similar to the linear eigenmodes in the slab ITG case, even exhibiting a critical gradient which arises due to resonant stabilisation, despite the low-dimensionality of the system. Unlike traditional gyro-fluid theories, which often involve an ad-hoc closure to capture these effects, the constrained optimal modes make no such approximation, providing a rigorous bound on fully kinetic linear instability growth.

\section{Linear Gyrokinetic Equation}
Here, we concern ourselves with the electrostatic limit of the gyrokinetic equation in flux tube geometry, where the fluctuations are periodic in the directions perpendicular to the magnetic field, which is given by $\mathbf{B} = \nabla \psi  \times \nabla \alpha $. In these coordinates, $\psi$ acts as a `radial' coordinate (in the sense that the plasma gradients will be in the coordinate $\psi$), and $\alpha$ is binormal to the magnetic field and the gradient direction. The coordinate along the magnetic field lines is denoted by $l$.  Because we wish to test the limits of the optimal modes' ability to bound the growth of linear instabilities, we must consider a scenario where the linear eigenmodes are well understood\footnote{In fact, as we will see later, we actually require \textit{complete} knowledge of the infinite set of linear modes to achieve our tightest bound on linear growth.}. To this end, we consider the case of a uniform magnetic geometry (henceforth called `slab' geometry), in which a single Fourier mode along the magnetic field line can be considered, such that $\partial/ \partial l \to i k_\parallel$. 

Moreover, we consider only a single kinetic species, focusing on the familiar limit of kinetic ions with adiabatic electrons \citep{biglariToroidalIonpressuregradientdrivenDrift1989, kadomtsevTurbulenceToroidalSystems1970, plunkCollisionlessMicroinstabilitiesStellarators2014}. 
In this scenario, the gyrokinetic equation becomes, 
\begin{equation}
    \frac{\partial g}{\partial t} + i v_{\|} k_\| g 
    = \frac{e F_{0}}{T_i}\left(\frac{\partial}{\partial t} + i\omega_{*}^T\right)\phi J_{0}.
    \label{eqn:GK_equation}
\end{equation}
Here, $g = g(\mu, E,\mathbf{k}_\perp, k_\|)$ is the (Fourier transformed) non adiabatic part of the perturbed ion distribution function. The phase space variables are the particle energy $E =m v^2/2$ and the magnetic moment $\mu = m v_\perp^2/(2B)$, where $v^2 = v_\perp^2  + v_\parallel^2$, where $m = m_i$ is the ion mass.
The ions are assumed to have a Maxwellian background distribution $F_{0} = n(\psi)/(v_{T}^3\pi^{3/2})\exp(-v^2/v_{T}^2)$ with a number density $n(\psi)$, where the temperature enters via the thermal speed $v_{T} = \sqrt{2T_i/m}$ and $T_i = T_i(\psi)$. The plasma gradients enter via the energy-dependent diamagnetic frequency $\omega_{*}^T = \omega_{*}\left(1 + \eta[v^2/v_{T}^2 - 3/2]\right)$ where $\omega_{*} = ({k_\alpha T_i}/{e})({\mathrm{d}\ln n}/{\mathrm{d}\psi})$ and $\eta = \mathrm{d} \ln T_i/ \mathrm{d}\ln n$. The Bessel function, which arises due to the gyro-average, has the argument $J_{0} = J_0(k_\perp v_\perp/\Omega)$ where $\Omega = e B/ m$. The system is closed by quasi-neutrality,
\begin{equation}
    \frac{e^2 n}{T_i}(1 + \tau) \phi = e   \int_{-\infty}^{\infty}\mathrm{d} v_\| \int_0^{\infty} \,\mathrm{d} v_\perp v_\perp  2 \pi  g J_{0} = e \int_{-\infty}^{\infty} \bar g \mathrm{d}v_\|,
    \label{eqn:QN}
\end{equation}
where $\phi$ is the Fourier transformed electrostatic potential and $\tau$ is the ion-electron temperature ratio $\tau = T_i/T_e$. Additionally, we have defined the reduced distribution function $\bar g(v_\|)= 2\pi \int_{0}^{\infty} v_\perp g J_0 \mathrm{d} v _\perp$.

Because the wave-particle resonance is one-dimensional in phase space in slab geometry, we may integrate Equation \eqref{eqn:GK_equation} over $v_\perp$ to express the gyrokinetic equation in terms of $\bar g$. This yields
\begin{equation}
    \frac{\partial \bar g}{\partial t} + i v_{\|} k_\| \bar g 
    = \frac{e \bar F_{0}}{T_i}\left[G_{\perp 0}\left(\frac{\partial}{\partial t} + i\omega_{*}(1 + \eta(x_\|^2 - 3/2 ))\right) + i \omega_* \eta G_{\perp 2}\right]\phi,
    \label{eqn:GK_equation_reduced_vspace}
\end{equation}
where $\bar F_0 = {n}/({v_T \sqrt{\pi}})e^{-v_\|^2/ v_T^2}$ and $x_\| = v_\|/v_T$. The functions $G_{\perp0}$ and $G_{\perp2}$ can be written in terms of the familiar $\Gamma_n(b) = I_n(b)e^{-b}$ functions of gyrokinetic theory where $I_n$ is a modified Bessel function and $b = k_\perp^2 \rho^2$ with $\rho = v_T/\Omega$ (see \cite{plunkEnergeticBoundsGyrokinetic2023} for specifics),
\begin{equation}
    G_{\perp 0}(b) = \Gamma_0(b),
\end{equation}
\begin{equation}
    G_{\perp 2}(b) = \Gamma_0(b) - b[\Gamma_0(b) - \Gamma_1(b)].
\end{equation}
The linear eigenmodes of \eqref{eqn:GK_equation_reduced_vspace} have been well studied \citep{kadomtsevTurbulenceToroidalSystems1970, plunkCollisionlessMicroinstabilitiesStellarators2014} allowing clear comparison between the linear growth rates and the bounds derived here.

\section{Tightest possible energetic bounds}
\label{subsection:CVK}
Before attempting to tailor the optimal mode analysis to bounding the growth of linear instabilities, we will determine whether there is a fundamental limitation of the optimal modes when it comes to bounding linear eigenmode growth. In other words, how tightly can the growth of the linear modes be bounded by the instantaneous growth of an energetic norm? To answer this question, we will leverage techniques that are seldom used in gyrokinetics \citep{heningerIntegralTransformTechnique2018, plunkLandauDampingTurbulent2013,plunkNonlinearStabilityQuasitwodimensional2015}.
We begin by defining $f = \bar g - e \bar F_0 G_{\perp 0} \phi/T_i$, allowing us to write Equation~\eqref{eqn:GK_equation_reduced_vspace} as, 
\begin{equation}
\label{eqn:GK_f_CVK}
    \frac{\partial f}{\partial t} + i v_\| k_\| f =  - i k_\| G(v_\|, k_\|, b)\int f \mathrm{d} v_\|
\end{equation}
where we have used quasi-neutrality to close the system in terms of an intergo-differential equation and defined,
\begin{equation}
    G(v_\|) = \frac{\bar F_0}{(1 + \tau - G_{\perp 0})}\left[v_\| G_{\perp 0} - \frac{\omega_*}{k_\|}\left(G_{\perp 0}(1 + \eta(x_\|^2 - 3/2)) + \eta\,  G_{\perp 2}\right) \right],
\end{equation}
where we have suppressed the dependence on other parameters e.g., $k_\|$ and $k_\perp$.
Equation~\eqref{eqn:GK_f_CVK} is equivalent to the linear system studied by \cite{plunkLandauDampingTurbulent2013} with the additional inclusion of finite-Larmor-radius effects. 

\subsection{Case--Van Kampen energy}
The linear eigenmodes of this equation satisfy,
\begin{equation}
    (\omega - v_{\|}k_\parallel )f 
    = k_\parallel G(v_\parallel) \int f(v_\parallel') \mathrm{d} v_\parallel',
    \label{eqn:slab_eiegnmodes}
\end{equation}
where $\omega = \omega_r + i \gamma$ is the eigenvalue. We may also set the normalisation of the solutions to,
\begin{equation}
    \int_{-\infty}^{\infty} f(v_\|) \mathrm{d}v_\| = 1.
\end{equation}
Equation \eqref{eqn:slab_eiegnmodes} admits, depending on the choice of parameters (i.e., $k_\parallel$, $\omega_*$, $\eta$ , $b$, and $\tau$), a discrete set of damped and unstable modes (see Appendix~\ref{appendix:lineardisp}) which come in conjugate symmetric pairs (due to the time reversal symmetry of the kinetic equation without collisions) which we denote as
\begin{equation}
f_n = G(v_\parallel)/(\omega_n/k_\parallel - v_\parallel),
\end{equation}
and a continuum of singular undamped `Van Kampen' modes with $\gamma = 0$ which we denote as,
\begin{equation}
f_\omega  = P[{G(v_\parallel})/{(\omega/k_\parallel - v_\parallel)}] + \lambda(\omega)\delta(\omega/k_\parallel - v_\parallel),    
\end{equation}
where $P$ symbolically indicates that the Cauchy principal value should be taken upon integration. Integrating \eqref{eqn:slab_eiegnmodes} for a discrete mode yields the linear dispersion relation for the slab ITG mode,
\begin{equation}
    1  - \int_{-\infty}^{\infty}   \frac{k_\parallel G(v_\parallel)}{\omega_n - v_\parallel k_\parallel} \mathrm{d} v_\parallel =  0. 
    \label{eqn:slab_mode_dispersion_relation}
\end{equation}
In the case of the continuum modes, integrating \eqref{eqn:slab_eiegnmodes} does not constitute a dispersion relation, due to the singularity at $v_\| = \omega /k_\|$, and instead sets the normalisation of the singular modes that determines $\lambda(\omega)$ \citep[see][]{van1967theoretical},
\begin{equation}
    \lambda(\omega)  =  1  - P \int \frac{k_\| G(v_\|)}{\omega - v_\| k_\|} \mathrm{d}v_\|.
\end{equation}
This system is directly analogous to the system studied by Case and Van Kampen, where it was shown by \cite{casePlasmaOscillations1959} that this set of linear eigenmodes, comprised of the discrete and continuum modes, is \textit{complete}, in the sense that any distribution function $f(v_\parallel,t )$ can be written as
\begin{equation}
    f(v_\parallel, t) = \sum_n a_n(t) f_n(v_\parallel)  + \int A(\omega, t) f_\omega(v_\parallel)\, \mathrm{d}\omega.
\end{equation}
The coefficients of this eigenmode decomposition can be computed by exploiting the orthogonality of eigenmodes with respect to the eigenmodes of the adjoint equation of Eqn.~\eqref{eqn:slab_eiegnmodes}, which can be derived by computing the adjoint of the linear operator associated with \eqref{eqn:slab_eiegnmodes}  under the $L^2$-norm in $v_\parallel$, and is given by,
\begin{equation*}
        \frac{\partial \tilde{f}}{\partial t} + i v_{\|}k_\parallel \tilde{f} 
    = - {i}k_\parallel \int G(v_\parallel') \tilde{f}(v_\parallel') \mathrm{d} v_\parallel'.
    \label{eqn:adjoint_problem}
\end{equation*}
The eigenmodes of the adjoint problem we denote as $\tilde{f}_\omega$ for the continuum branch and $\tilde{f}_n$ for the discrete branch. As shown in \cite{casePlasmaOscillations1959}, the eigenmodes of \eqref{eqn:slab_eiegnmodes} and the adjoint problem satisfy the orthogonality relations
\begin{equation}
    \int_{-\infty}^{\infty}\tilde{f}_n(v_\parallel) f_m(v_\parallel)\, \mathrm{d}v_\parallel = \delta_{n m}C_n
\end{equation}
 and 
\begin{equation}
    \int_{-\infty}^{\infty}\tilde{f}_{\omega'}(v_\parallel) f_{\omega}
    (v_\parallel)\, \mathrm{d}v_\parallel = \delta(\omega - \omega')C_\omega
\end{equation}
for the discrete and continuum branches, respectively. By leveraging these expressions we can project any distribution function onto the complete basis of eigenmodes. We now wish to use these features of the kinetic equation to construct an energetic norm which is `aware' of the linear eigenmodes in which we are interested.
 
To do this, we first project \eqref{eqn:GK_f_CVK} onto the basis of discrete eigenmodes by applying,
 \begin{equation}
       \frac{1}{C_n}\int_{-\infty}^{\infty} \tilde f_n(v_\parallel')(\ldots) \mathrm{d}v_\parallel'.
 \end{equation}
For the discrete spectrum, the adjoint modes have velocity space dependence $\tilde{f}_{n} = 1/(\omega_n/k_\parallel - v_\parallel)$. Applying this operator to \eqref{eqn:GK_f_CVK} and noting that $\omega_n$ must satisfy \eqref{eqn:slab_mode_dispersion_relation} yields the simple relation,
 \begin{equation}
     \frac{\partial a_n}{\partial t} = -i \omega_n a_n.
 \end{equation}
 This implies that
 \begin{equation}
     \frac{\mathrm{d}}{\mathrm{d} t} \sum_n |a_n|^2 = 2 \sum_n \gamma_n |a_n|^2,
     \label{eqn:discrete_amplitudes}
 \end{equation}
where we have summed over all the discrete modes. Equation~\eqref{eqn:discrete_amplitudes} states that the amplitudes of the projection coefficients all grow in time according to their growth rate $\gamma_i$. Next, we turn our attention to the continuum modes. Where the (singular) adjoint modes are of the form $\tilde{f}_\omega = P[{1}/{(\omega/k_\parallel - v_\parallel)}] + \tilde\lambda(\omega)\delta(\omega/k_\parallel - v_\parallel)$. Performing the projection in a similar manner as for the discrete spectrum, noting that the adjoint modes satisfy $\tilde\lambda(\omega) = \lambda(\omega)/G(\omega/k_\parallel)$ and recognising the normalisation condition of the Van Kampen modes gives, $\partial  A(\omega ,t) /{\partial t }= i \omega A(\omega ,t)$. Note that this projection onto the continuum is also equivalent to the Morrison $G$-transform which has been studied in detail \citep{morrisonDielectricEnergyPlasma1992, heningerIntegralTransformTechnique2018, plunkLandauDampingTurbulent2013}. Because $\omega$ is purely real for these modes, we have,
\begin{equation}
\label{eqn:contuinuum_amplitudes}
    \frac{ \mathrm{d}}{\mathrm{d} t }\int |A(\omega, t)|^2 \mathrm{d} \omega  = 0.
\end{equation}
Now by combining \eqref{eqn:discrete_amplitudes} and \eqref{eqn:contuinuum_amplitudes} we can construct what we will refer to as the Case--Van Kampen energy, which we define as,
\begin{equation}
    E = \sum_n |a_n(t)|^2 + \int |A(\omega, t)|^2 \mathrm{d} \omega. 
\end{equation}
The energy balance equation for $E$ is now simply,
\begin{equation}
\label{eqn:energy_balance}
    \frac{\mathrm{d} E}{\mathrm{d} t} = 2 \sum_n \gamma_n |a_n|^2.
\end{equation}
The energy $E$ constitutes a positive definite norm of any distribution function $f(v_\parallel)$ due to the completeness of the eigenmodes. For a given $f$, E grows (or damps) at a rate determined by the magnitude of its projection onto the discrete eigenmodes.

\subsection{Optimal modes and tightest possible bounds}
The Case--Van Kampen energy is a natural choice as a norm for bounding the growth of linear instabilities as it is simply a sum of eigenmode amplitudes, thus it exhibits no `non-normal' or `transient growth' in the the absence of linear instability. While such transient growth is important for understanding nonlinear instability growth e.g., in scenarios where subcritical turbulence is of interest \citep{landremanGeneralizedUniversalInstability2015,barnesTurbulentTransportTokamak2011,wykTransitionSubcriticalTurbulence2016,krommesSubmarginalProfilesTurbulent1997}, it seems relatively unimportant in scenarios in which the relevant parameters are well above the linear instability threshold where robustly growing instabilities can be expected. Note that $E$ also does not decay if the system undergoes Landau damping, unlike some other norms e.g, $|\phi|^2$ \citep{landauElectronPlasmaOscillations1946}.

We can now derive the optimal modes of $E$ by writing the energy balance \eqref{eqn:energy_balance}  as $\rm{d} E/ \rm{d} t  = 2K$ and defining the instantaneous growth rate of $E$ as $\Lambda \equiv K/E$. The optimal modes are the solutions $f$ which extremise $\Lambda$, and can be constructed by writing $E \equiv (f, \mathcal{E} f)$, and $K \equiv (f, \mathcal{K} f)$ with the familiar inner product,
\begin{equation}
    (f_1, f_2) = \int \frac{f_1^* f_2}{\bar F_0} \mathrm{d} v_\parallel.
\end{equation}
The optimal modes then satisfy the generalised eigenvalue problem, 
\begin{equation}
    \label{eqn:CVK_eigenvalue_problem}
    \Lambda \operatorname{\mathcal{E}} f = \operatorname{\mathcal{K}} f
\end{equation}
where the operators $\mathcal{E}$ and $\mathcal{K}$ are given by
\begin{equation}
    \operatorname{\mathcal{E}}f = \bar F_0\left(\sum_n \frac{\tilde f_n^*}{| C_n |^2}\int \tilde f_n' f' \operatorname{d}v_\parallel'  + \int \mathrm{d}\omega  \frac{\tilde f_\omega^*}{| C_\omega |^2} \int \tilde f_\omega' f' \operatorname{d}v_\parallel'\right)
\end{equation}
and,
\begin{equation}
    \operatorname{\mathcal{K}}f = \bar F_0\sum_n \gamma_n \frac{\tilde f_n^*}{| C_n |^2}\int \tilde f_n' f' \operatorname{d}v_\parallel'.
\end{equation}
To solve the optimal mode problem we now consider the projection of \eqref{eqn:CVK_eigenvalue_problem} with a discrete eigenmode of the linear problem $f_m$ with the inner product $\Lambda (f_m, \mathcal{E} f) = (f_m, \mathcal{K} f)$. Noting the orthogonality of the eigenmodes with respect to the adjoint eigenmodes yields,
\begin{equation}
    \Lambda \sum_n \frac{\delta_{n,m} C_m^*}{|C_n|^2}\int \tilde{f}_n' f' \mathrm{d}v_\parallel'  = \sum_n \gamma_n \frac{\delta_{n,m} C_m^*}{|C_n|^2}\int \tilde{f}_n' f' \mathrm{d}v_\parallel' ,
\end{equation}
giving,
\begin{equation}
    \Lambda = \gamma_n.
\end{equation} Similarly, considering the projection with a continuum mode yields $\Lambda = 0$. Thus there is a one-to-one correspondence between the linear growth spectrum and the optimal $\Lambda$ solutions. Moreover, the optimal modes maximising $E^{-1}\mathrm{d}E/ \mathrm{d} t$ are exactly the linear eigenmodes.

This may feel like a somewhat trivial consequence of the diagonalisation of the linear operator by the linear eigenmodes, but there are several takeaways from this result. Firstly, the growth of linear eigenmodes can be bounded with arbitrary tightness by the optimal modes of an energetic norm, with the tightest possible bound (i.e., equality of \eqref{eqn:gamma_lambda_ineq}) being given by the Case--Van Kampen energy. We note that this tightest bound may not be given uniquely by this energy norm.  For instance, $E$ can be defined with arbitrary weights on each of the amplitude coefficients. 

Secondly, any kinetic system for which the Case--Van Kampen energy is a nonlinear invariant is nonlinearly stable if it is linearly stable because $\Lambda$ will be identically zero for all $k_\perp$, i.e., subcritical turbulence is impossible. An example of such a system is the 2D drift-kinetic toroidal ITG discussed by \cite{plunkNonlinearStabilityQuasitwodimensional2015}. As presented in \cite{delsoleNecessityInstantaneousOptimals2004, landremanGeneralizedUniversalInstability2015, plunkEnergeticBoundsGyrokinetic2022}, for statistically steady turbulence to exist in these systems where $E$ is a nonlinear invariant,  it must be the case that $\sum_k \langle \mathrm{d} E/ \mathrm{d} t \rangle = 0 $ where $\langle \ldots \rangle$ denotes a time average over the turbulent time scale. In this case $\sum_{n,k} \gamma_{n,k} \langle|a_{n,k}|^2\rangle = 0$, such that, in the turbulent phase, the distribution function must project equally on to growing and damped eigenmodes on average. 

\section{Constrained optimal modes}
We have demonstrated that linear instability growth may be bounded with arbitrary tightness by the instantaneous optimals of an energetic norm. However, these tightest bounds require complete knowledge of the linear spectrum. In this section, we demonstrate that we can construct energetic bounds that have the same qualitative dependence on key parameters as the linear growth rate by including only some features of the linear eigenmodes in the analysis. This avoids the requirement of prior knowledge of the solution to the linear problem. 

A key feature of the linear eigenmodes, which has been absent from the optimal mode analyses involving simple energetic norms, is the real frequency $\omega_r$. This oscillation frequency determines the relative phases of the different degrees of freedom of the eigenmode (e.g., the various fluid moments) and is associated with some of the stabilising `resonant' effects that are not captured by the instantaneous upper bounds, which do not have an associated real frequency. As such, the optimal modes considered thus far, with the exception of the Case--Van Kampen optimals discussed in the previous section, are largely unconstrained in the allowable phase relations between their different degrees of freedom.

Here, we seek a means to include phase information into the optimal mode analysis of a simple, positive definite energy norm --- the Helmholtz free energy \citep{helanderEnergeticBoundsGyrokinetic2022}. For the reduced-dimensionality equation \eqref{eqn:GK_equation_reduced_vspace}, Helmholtz free energy balance is,
\begin{equation}
\label{eqn:freeenergybalance}
    \frac{\mathrm{d} H}{\mathrm{d} t}  = 2 D,
\end{equation}
where 
\begin{equation}
    H = T_i \int \frac{| \bar g|^2}{\bar F_0}\mathrm{d} v_\parallel  - \frac{e^2 n}{T_i}(1 + \tau)G_{\perp 0} |\phi|^2,
\end{equation}
and
\begin{equation}
   D = \mathrm{Re}\left\{ i  \omega_* \eta\,  G_{\perp 0}  e\phi \int \bar g^*x_\|^2 \,\mathrm{d} v_\| \right\}.
\end{equation}
To include the frequency information, we \textit{constrain} the optimal mode analysis, such that the allowed distribution functions are those that share some key features with the linear modes. Thus, we seek distribution functions that maximise $\mathrm{d}H/ \mathrm{d} t$ subject to a set of constraint equations that are also satisfied by the linear eigenmodes. 

Because the free energy extraction $D$ depends only on two `fluid' moments of $\bar g$, a natural choice for these constraints are the linear moment equations. To keep the analysis relatively simple, we consider only the two lowest-order moments as constraints. We denote these moments of $\bar g $ by,
\begin{equation}
    \kappa_1  = \frac{1}{n}\int \bar g \mathrm{d} v_\|,
\end{equation}
\begin{equation}
        \kappa_2  = \frac{1}{n}\int \left(\frac{v_\|}{v_T }\right)\bar g \, \mathrm{d} v_\|,
\end{equation}
\begin{equation}
    \kappa_3  = \frac{1}{n}\int \left(\frac{v_\|^2}{v_T^2 }\right)\bar g \, \mathrm{d} v_\|.
\end{equation}
The moment equations for the density, $\kappa_1$, and the parallel flow, $\kappa_2$, obtained by taking the respective moments of \eqref{eqn:GK_equation_reduced_vspace}, are given by,
\begin{equation}
    \frac{\partial \kappa_1}{\partial t} + i v_T k_\| \kappa_2  = \frac{1}{1 + \tau}\left[G_{\perp 0}\left(\frac{\partial}{\partial t} + i\omega_{*}(1 - \eta)\right) + i \omega_* \eta G_{\perp 2}\right]\kappa_1
\end{equation}
and,
\begin{equation}
    \frac{\partial \kappa_2}{\partial t} + i v_T k_\| \kappa_3 = 0.
\end{equation}
We now require that these moments evolve in time like linear eigenmodes (i.e., they satisfy the same phase relation as the moments of a linear eigenmode), such that $\partial \kappa_{1,2}/\partial t = -i \omega' \kappa_{1,2}$, where $\omega' = \omega_r' + i \gamma'$. The linear moment equations then become, 
\begin{equation}
\label{eqn:density_moment_constraint}
    \alpha \kappa_1  - v_T k_\| \kappa_2 = 0
\end{equation}
and
\begin{equation}
\label{eqn:flow_moment_constraint}
    \omega' \kappa_2  - v_T k_\| \kappa_3 = 0,
\end{equation}
where we have defined,
\begin{equation}
    \alpha = \omega' - \frac{1}{1 + \tau }\bigg(G_{\perp 0}\left[\omega' - \omega_*(1 - \eta)\right] - \omega_*\eta\,  G_{\perp 2} \bigg).
\end{equation} 
In addition to these phase relations between moments, linear eigenmodes have the property that free energy increases at exactly twice the growth rate of the mode, which can be expressed in terms of the free energy balance equation \eqref{eqn:freeenergybalance} as an additional constraint
\begin{equation}
\label{eqn:freeenergyconstraint}
    D = \gamma' H.
\end{equation}
The distribution functions that satisfy equations \eqref{eqn:density_moment_constraint},  \eqref{eqn:flow_moment_constraint} and \eqref{eqn:freeenergyconstraint} make up a subspace of the larger space of all possible $\bar g$ and, importantly, \textit{the linear eigenmodes are themselves included in that subspace}. In other words, $\omega'$ need not be an eigenvalue of \eqref{eqn:GK_equation_reduced_vspace}, but the true eigenmodes also satisfy \eqref{eqn:density_moment_constraint},  \eqref{eqn:flow_moment_constraint} and \eqref{eqn:freeenergyconstraint} with $\omega' \to \omega $.  

Therefore, distribution functions which extremise free-energy growth, while satisfying the constraints \eqref{eqn:density_moment_constraint},  \eqref{eqn:flow_moment_constraint} and \eqref{eqn:freeenergyconstraint} place a rigorous upper bound on the growth rate of true eigenmodes.  To solve this problem formally, we consider the Lagrangian,

\begin{multline}
        L \equiv D - \Lambda(H - H_0) - \lambda_1(\gamma' H - D) \\- \lambda_2^*(\alpha^* \kappa_1^* - v_T k_\| \kappa_2^*)   - \lambda_2(\alpha \kappa_1 - v_T k_\| \kappa_2) \\ - \lambda_3^*(\omega'^* \kappa_2^* - v_T k_\| \kappa_3^*) - \lambda_3(\omega' \kappa_2 - v_T k_\| \kappa_3).   
\end{multline}
Here, to extremise $L$ is to extremise the free-energy drive $D$, subject to several constraints, which are enforced by Lagrange multipliers denoted by $\Lambda$ (which will become the optimal growth) and $\lambda_n$. Here, $\Lambda$ keeps the free energy fixed to some normalising value $H_0$, the complex multipliers $\lambda_2$ and $\lambda_3$ enforce the moment constraints and $\lambda_1$ ensures that the value of $\gamma'$ from these moment constraints is consistent with free energy balance. Each of these constraints are enforced at the extrema of $L$, as is evident from taking $\delta L/ \delta \Lambda =0$ and $\delta L/ \delta \lambda_n =0$.

Computing the optimal modes via $\delta L/ \delta \bar g =0$ and making use of the inner product,
\begin{equation}
    (\bar g_1, \bar g_2 ) = T_i \int \frac{\bar g_1^* \bar g_2}{\bar F_0}\mathrm{d} v_\|, 
\end{equation} 
we arrive at the following kinetic problem for the optimal $\bar g$,
\begin{multline}
\label{eqn:constrianed_problem}
    (\Lambda - \gamma' \lambda_1)\left(\bar g  - \frac{\bar F_0 G_{\perp0}}{(1 + \tau)}\kappa_1 \right) = (1 - \lambda_1) \frac{i \omega_* \eta \bar F_0 G_{\perp 0}}{2(1 +\tau)}(x_\|^2 \kappa_1 - \kappa_3)\\
    - \frac{\lambda_2^* \bar F_0}{n T_i}(\alpha^* - x_\| v_T k_\|)
    - \frac{\lambda_3^*}{n T_i} \bar F_0(\omega'^*x_\| - x_\|^2 v_T k_\|).
\end{multline}
This problem must be solved subject to constraint equations \eqref{eqn:density_moment_constraint}, \eqref{eqn:flow_moment_constraint} and \eqref{eqn:freeenergyconstraint}.

Upon enforcing the constraint $\gamma' H = D$ in the above expression, we find the expected result that $\Lambda = \gamma'$, such that the Lagrange multiplier once again has the physical interpretation of the linear growth rate. We see that after enforcing this, the $\lambda_1$ Lagrange multiplier is arbitrary (as long as it is non-zero) such that it amounts to a renormalisation of the other multipliers. The remaining unknowns can be eliminated by noting that the system is closed by projecting onto the $\kappa_n$ moments. The system of moment equations which results from this projection is given in Appendix \ref{appendix:systemofeqns}.  After reducing the system of equations, we arrive at a
quintic polynomial for $\tilde\Lambda = \Lambda/({\omega_*\eta})$, for which $\tilde\Lambda = 0$ is a solution (corresponding to the null-space of $D$). The remaining solutions are given by the quartic equation,
\begin{equation}
\label{eqn:quartic}
    P \tilde\Lambda^4  + Q \tilde\Lambda^2  + R = 0,
\end{equation}
where the coefficients $P$, $Q$ and $R$ are given in appendix \ref{appendix:coefficients}. We note that only solutions to \eqref{eqn:quartic} that yield a real value of $\Lambda$ are valid. Complex solutions, which are inconsistent with \eqref{eqn:freeenergyconstraint}, are discarded. Solving this equation yields and optimal mode growth rate $\Lambda$, with an explicit dependence on the parameter $\omega_r'$. The upper bound on linear growth is given by
\begin{equation*}
    \Lambda_{\mathrm{max}}\equiv \max\limits_{\omega_r'} \Lambda. 
\end{equation*}
The maximising value of $\omega_r'$ can, in principle, be found analytically. However, it is generally more convenient to evaluate numerically or with a computer algebra program like \verb|Mathematica|. 

\subsection{Features of the constrained bound}

\begin{figure}
    \centering
\includegraphics[width=0.6\linewidth]{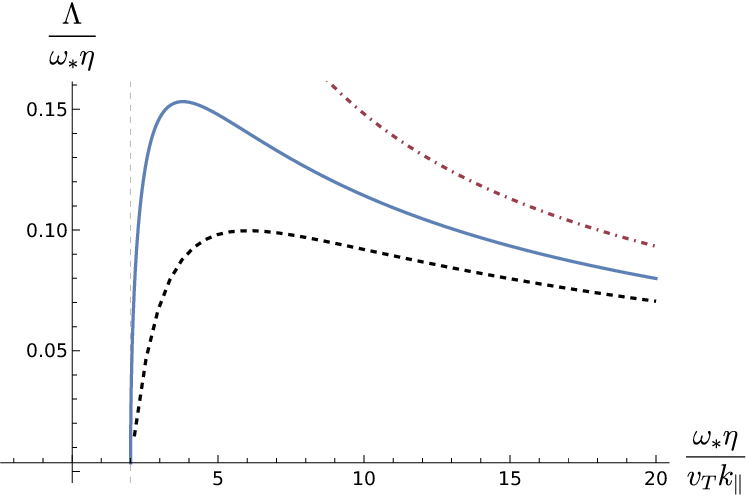}
    \caption{Upper bound on linear growth (blue) alongside the linear growth rate (dashed black) from the linear dispersion relation \eqref{eqn:lineardispersion}, plotted as a function of the instability parameter $\kappa_\|$ in the low-$k_\perp$ limit with $\eta\to \infty$ and $\tau =1$. Also shown is the fluid-limit dispersion relation (dot-dashed red) \citep{plunkCollisionlessMicroinstabilitiesStellarators2014}, which diverges as $\kappa_\| \to 0$. The vertical light-grey dashed line is the critical gradient of \cite{kadomtsevTurbulenceToroidalSystems1970}.}
    \label{fig:low-k-result}
\end{figure}
To gain insight into the behaviour of the solutions of \eqref{eqn:quartic}, we now consider some simple limits and explore the general features of the constrained optimal mode growth.
\subsubsection{Resonant stabilisation}
Firstly, to facilitate comparison with the results of \cite{plunkEnergeticBoundsGyrokinetic2023} we consider the long-wavelength limit where $b \to 0$ and consider a scenario without a density gradient where $\eta \to \infty$. In this limit, we have $G_{\perp 0,2} \to 1$.

In Figure~\ref{fig:low-k-result}, we show the behaviour of the upper bound alongside the solution of the linear dispersion relation as a function of the instability parameter $\kappa_\| = \omega_*\eta/(v_T k_\|)$. As one would expect, the bound agrees well with the linear growth rates for  large $\kappa_\|$, where the linear eigenmodes become more fluid-like as the lowest moments dominate the fluid hierarchy. In contrast to previous bounds \citep{plunkEnergeticBoundsGyrokinetic2023}, the constrained bound for the slab ITG exhibits a maximum growth rate at a finite $\kappa_\|$, as is the case for linear eigenmodes. This also in contrast to the simple non-resonant dispersion relation for the slab ITG mode shown in Fig.~\ref{fig:low-k-result}, which actually diverges in this limit \citep{plunkCollisionlessMicroinstabilitiesStellarators2014},
\begin{equation}
    \frac{\gamma}{\omega_*\eta} = (2 \tau \kappa_\|)^{-\frac{1}{3}}.
\end{equation}

Remarkably, the constrained bound also exhibits a \textit{critical gradient}, a value of $\kappa_\|$ below which only $\Lambda = 0$ is a valid solution. The criterion for this critical value of $\kappa_\|$ can be found by setting $R = 0$, such that $\Lambda = 0$ is a repeated root of \eqref{eqn:quartic} and seeking the lowest positive value of $\kappa_\|$ for which this can occur. In this limit, we have $\kappa_{\|, \mathrm{cr}}  = \sqrt{2\tau (1 + \tau)}$, which is precisely the critical gradient which can be derived from the linear slab ITG dispersion relation \citep{kadomtsevTurbulenceToroidalSystems1970, plunkCollisionlessMicroinstabilitiesStellarators2014}. Indeed, we see agreement between the critical gradient of the linear eigenmode and the bound in Figure~\ref{fig:low-k-result}. This agreement in the ``resonant'' limit of low-$\kappa_\|$ is notable given that the most unstable linear eigenmode becomes increasingly singular as the critical $\kappa_\|$ is approached. As a result, one might not expect to capture this behaviour with a finite-dimensional system of moment equations. On the other hand, the critical gradient expression derived by \cite{kadomtsevTurbulenceToroidalSystems1970}, which amounts to solving a two-dimensional system of equations, gives some indication that the resonant stabilisation of the slab ITG is not as complicated as one may expect. The constrained optimal modes are, evidently, successful in capturing this resonant stabilization due to the inclusion of information about the real frequency via $\omega_r'$, which is responsible for the effect in the linear dispersion relation.

\subsubsection{Density gradient stabilisation}
\begin{figure}
    \centering
    \includegraphics[width=0.6\linewidth]{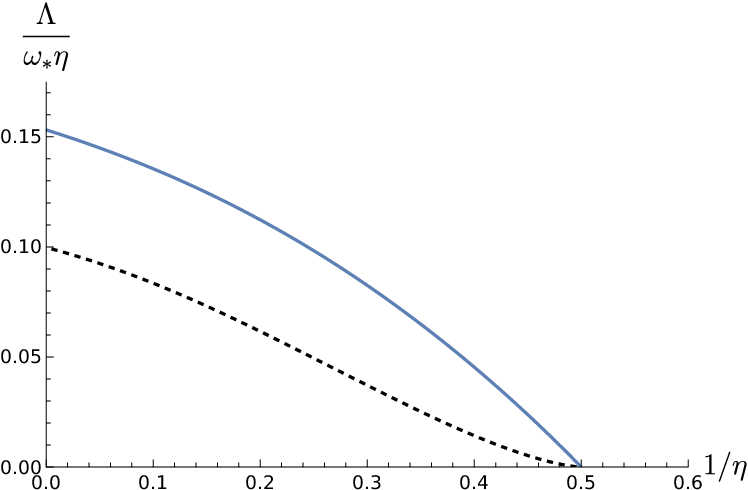}
    \caption{Upper bound on linear growth (blue) alongside the linear growth rate (dashed black) from the linear dispersion relation \eqref{eqn:lineardispersion}, plotted as a function of $1/ \eta$ in the low-$k_\perp$ limit with $\tau = 1$. Here, both the upper bound and the linear growth rate have been maximised over $k_\|$. }
    \label{fig:eta_scan}
\end{figure}

We now consider a finite value of $\eta$, such that the density gradient is non-zero. This is known to have a significant stabilising effect on linear ITG instabilities. The previous, nonlinear bounds of \cite{helanderEnergeticBoundsGyrokinetic2022, plunkEnergeticBoundsGyrokinetic2023} do not capture this effect because the adiabaticity of the electrons precludes the density gradient from being a source of free energy. In Figure~\ref{fig:eta_scan}, we show that the constrained optimal growth rate has the same qualitative dependence on $\eta$ as the linear instability growth. Once again, we see that the critical value of $\eta$ agrees with that of the linear dispersion relation. In the linear theory, this stabilisation by the density gradient is connected with the real frequency of the ITG, hence why the constrained upper bound captures this effect.

\subsubsection{Finite-Larmor-radius effects}

\begin{figure}
    \centering
    \includegraphics[width=0.6\linewidth]{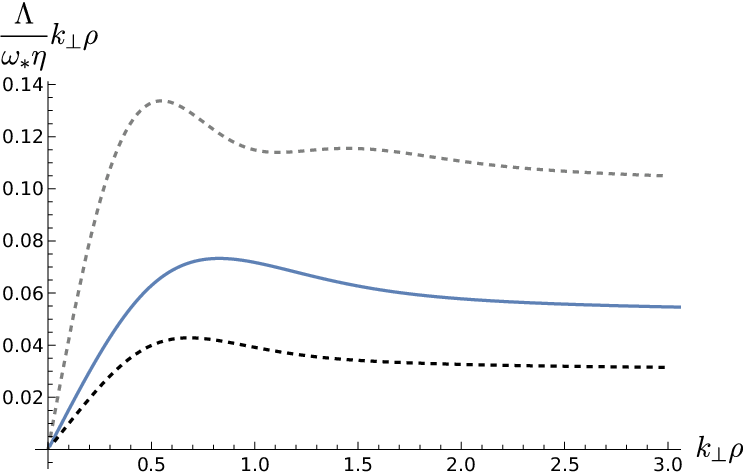}
    \caption{Upper bound on linear growth (blue) alongside the linear growth rate (dashed black) from the linear dispersion relation \eqref{eqn:lineardispersion} and the nonlinear bound of \cite{helanderEnergeticBoundsGyrokinetic2022, plunkEnergeticBoundsGyrokinetic2023} (dashed grey), plotted as a function of $k_\perp\rho$ in the limit of $\eta \to \infty$ for $\tau = 1$, where both the constrained upper bound and the linear growth rate have been maximised over $k_\|$.}
    \label{fig:FLR}
\end{figure}

Finally, we consider how the upper bound from the constrained optimal modes depends on the normalized perpendicular wavenumber $k_\perp \rho$. In the case of the slab ITG, because the resonance does not involve the perpendicular velocity, the Finite-Larmor-Radius (FLR) effects manifest via the velocity-space-independent functions $G_{\perp1,2}$. As a result, the most consequential step for capturing the FLR dependence of the slab ITG is the integration over the $v_\perp$ coordinate when going from \eqref{eqn:GK_equation} to \eqref{eqn:GK_equation_reduced_vspace}. In Figure~\ref{fig:FLR}, we see that the constrained optimal growth has a very similar qualitative dependence on $k_\perp\rho$ to the linear growth rate, and is significantly lower than the pervious bound of \cite{plunkEnergeticBoundsGyrokinetic2023}.

\section{Conclusions}
Two pieces of theoretical progress towards tighter energetic bounds on linear instabilities have been made in this work. The first is the development of the Case--Van Kampen energy, which provides proof that linear energetic bounds can be made arbitrarily tight to the growth rate of linear eigenmodes. This special energy is of largely theoretical interest, requiring complete knowledge of the linear eigenspectrum to construct. Nevertheless, it provides a solid foundation for the endeavour of tightening linear bounds, as well as some indication of how such tighter bounds can be constructed, namely by including information from the linear spectrum. It remains to be seen if such an energy can be constructed in non-trivial magnetic geometry.

The second theoretical development made here was to consider the constrained optimal modes. These modes extremise the growth of the Helmholtz free energy subject to linear moment constraints and capture the key parametric dependencies of the linear growth rate. They are derived from a simple, thermodynamic energy norm without requiring an explicit solution to the linear problem, in keeping with the spirit of the original nonlinear bounds of Helander and Plunk. To compute the upper bound from these modes in the slab ITG limit requires solving a simple polynomial and optimising over a free parameter $\omega_r'$, much like the generalised free energy bounds \citep{plunkEnergeticBoundsGyrokinetic2023, costelloEnergeticBoundsGyrokinetic2024}. The inclusion of this real frequency in the optimal mode analysis introduces some key ``resonant'' features of the linear slab ITG; a critical gradient below which eigenmode growth is impossible, and density gradient stabilisation. The constrained optimals are similar to work by \cite{kotschenreutherTransportBarriersMagnetized2024}, where free energy growth of linear modes is dynamically constrained.

The constrained optimal mode theory developed here warrants comparison to traditional linear gyrofluid theory, wherein the infinite hierarchy of fluid moments (discussed in terms of a projection onto a Hermite-Laguerre basis in more modern work) is truncated to give an approximation of the linear eigenvalue problem. Such theories involve choosing a closure in the truncation of the moment system. This closure can be chosen in a myriad of ways but is often selected to capture resonant effects that are beyond the typical regimes of validity of a truncation based on the smallness of  higher order moments due to some asymptotic argument, e.g. collisional \citep{braginskiiTransportProcessesPlasma1965}  or strongly-driven closures \citep{plunkCollisionlessMicroinstabilitiesStellarators2014}. An optimal choice for the closure model is an open, and active, research problem, with proposed closures ranging from ad-hoc (but effective) closures \citep{hammettFluidMomentModels1990} to recently developed machine learning approaches \citep{huangMachinelearningHeatFlux2025, barbourMachinelearningClosureVlasovPoisson2025}.
In contrast, the constrained optimal modes capture the qualitative behaviour of the linear growth rate in the resonant limit despite consisting of only a few low-order moments of the distribution function, which close as a natural consequence of the variational form of the problem. Thus, the upper bound is equally valid (but not necessary as close to the linear growth rate) at all plasma parameters.

We note that the constrained optimal modes considered here were only constrained by two fluid moments equations but, in principal, any number can be included, for instance by projection onto the Hermite-Laguerre basis (see \cite{mandellLaguerreHermitePseudospectral2018} for details). The infinite set of linear constraints that would result from this projection would yield a constrained optimal mode with $\Lambda = \gamma_{\mathrm{max}}$ where $\gamma_\mathrm{max}$ is the largest linear growth rate because the optimisation subspace would be exactly the set of linear eigenmodes, which are the only distribution functions that simultaneously satisfy the infinite hierarchy of constraints. Thus, one would expect a bound that may be made progressively tighter to the linear eigenmode growth rate by adding more of these fluid constraints. 
This observation could form the basis of a numerical implementation of constrained optimal modes with an arbitrary number of fluid constraints, similar to the recently developed pseudo-spectral gyrokinetic codes \citep{freiMomentbasedApproachFluxtube2023, mandellGXGPUnativeGyrokinetic2024}.

The success of the constrained optimal modes in giving a computationally efficient upper bound of linear eigenmode growth warrants further investigation. Namely, this calculation should be extended to treat general magnetic geometry. If the performance of these bounds in the slab geometry are indicative of their performance in more general scenarios, they could be a powerful tool for applications like stellarator optimisation. The manifestation of a critical ITG temperature gradient is of particular interest, as optimisation schemes based on this principal have been demonstrated to be effective \citep{roberg-clarkCriticalGradientTurbulence2023}.


\section*{Funding}
This work has been carried out within the framework of the EUROfusion Consortium, funded by the European Union via the Euratom Research and Training Programme (Grant Agreement No 101052200 -- EUROfusion). Views and opinions expressed are however those of the author(s) only and do not necessarily reflect those of the European Union or the European Commission. Neither the European Union nor the European Commission can be held responsible for them. P.C. is supported by the Simons Foundation (Grant No. 560651).

\section*{Declaration of interests}
The authors report no conflict of interest.

\appendix
\section{Linear dispersion relation}
\label{appendix:lineardisp}
The dispersion relation for the discrete eigenmodes of \eqref{eqn:CVK_eigenvalue_problem} is given by,
\begin{equation}
\label{eqn:lineardispersion}
    1 + \tau + G_{\perp 0}\left(\xi Z(\xi) + \kappa_\| \left[\left(\frac{3}{2} - \frac{1}{\eta} \right)Z(\xi) - \xi(1 + \xi Z(\xi) \right] \right) - \kappa_\|G_{\perp2} Z(\xi) = 0,
\end{equation}
where $\xi = \omega/(v_T k_\|)$ and $Z$ is the plasma dispersion function,
\begin{equation}
    Z(\xi) = \frac{1}{\sqrt{\pi}}\int_{-\infty}^{\infty} \frac{e^{-t^2}}{t - \xi} \mathrm{d}t, 
\end{equation}
which is defined for $\mathrm{Im}(\xi) > 0$.
The roots of \eqref{eqn:lineardispersion} can be found numerically.  

\section{System of moment equations}
\label{appendix:systemofeqns}
By noting the following integrals, 
\begin{equation}
    \frac{1}{n}\int \bar F_0 \mathrm{d} v_\| = 1,
\end{equation}
\begin{equation}
    \frac{1}{n}\int \left(\frac{v_\|^2}{v_T^2}\right) \bar F_0 \mathrm{d} v_\| = \frac{1}{2},
\end{equation}
\begin{equation}
    \frac{1}{n}\int \left(\frac{v_\|^4}{v_T^4}\right)\bar F_0 \mathrm{d} v_\| = \frac{3}{4},
\end{equation}
Equation~\eqref{eqn:constrianed_problem} can be closed in terms of the $\kappa_n$ moments. The system of equations is as follows:
\begin{equation}
    \tilde\Lambda\left( 1 - \frac{G_{\perp 0}}{(1 + \tau)}\right)\kappa_1 = \frac{i}{2} \frac{G_{\perp0}}{1 + \tau}\left(\frac{\kappa_1}{2} - \kappa_3 \right)- \tilde{\lambda}^*_2\tilde{\alpha}^* + \tilde\lambda_3^* \frac{\kappa_\|^{-1}}{2},
\end{equation}
\begin{equation}
    \tilde\Lambda\left( \kappa_3 - \frac{G_{\perp 0}}{2(1 + \tau)}\kappa_1\right) = \frac{i}{2} \frac{G_{\perp0}}{1 + \tau}\left(\frac{3\kappa_1}{4} - \frac{\kappa_3}{2} \right)- \tilde{\lambda}^*_2\frac{\tilde{\alpha}^*}{2} + \tilde\lambda_3^* \frac{3\kappa_\|^{-1}}{4},
\end{equation}

\begin{equation}
    2\tilde\Lambda \kappa_2 = \tilde\lambda_2^*\kappa_\|^{-1} - \tilde\lambda_3^*\tilde{\omega}'^*,
\end{equation}
\begin{equation}
    \tilde{\alpha} \kappa_1  - \kappa_\|^{-1} \kappa_2 = 0,
\end{equation}
\begin{equation}
    \tilde{\omega}' \kappa_2  - \kappa_\|^{-1} \kappa_3 = 0,
\end{equation}
where we have repeated the constraint equations for clarity. We have defined $\kappa_\| = \omega_*\eta/(v_T k_\|)$, $\tilde \alpha = \alpha/ (\omega_*\eta)$, $\tilde\omega' =\omega'/ (\omega_*\eta) =  \tilde\omega_r' + i \tilde{\gamma}'$ and $\tilde{\lambda}_{3,4} = \lambda_{2,3 }/(n T_i \omega_*\eta (1 - \lambda_1))$.

\section{Coefficients}
\label{appendix:coefficients}
The coefficients of~\eqref{eqn:quartic} are given by,
\begin{equation}
    P = 4 \eta^2 \kappa_\|^4(1 + \tau - G_{\perp 0})^2
\end{equation}
\begin{multline}
        Q = 4 \eta^2 \kappa_\|^2(1 + \tau - G_{\perp 0})(2(1 +\tau) - G_{\perp 0}) + 4 \kappa_\|^4((G_{\perp 0}(\eta - 1) + G_{\perp 2})^2 \\+ 2 \eta(1 + \tau - G_{\perp 0})(G_{\perp 0}(\eta  - 1) - G_{\perp2} \eta)\omega_r  + 2 \eta^2 (1 + \tau  - G_{\perp 0})^2 \omega_r^2)
\end{multline}
\begin{multline}
    R = 4 \kappa_\|^4 \omega_r^2( G_{\perp 0}(1 - \eta(1 + \omega_r)) + \eta(G_{\perp 2} +(1 + \tau) \omega_r ))^2 \\
    + \kappa_\|^2(4G_{\perp2}\eta^2(G_{\perp2} + \omega_r(1 + \tau)) - 4 G_{\perp 0}\eta(1 + \tau)\omega_r(\eta(2 + \omega_r) - 1) \\
    + 2 G_{\perp 0} G_{\perp 2} \eta(4(\eta \omega_r -1) + 5\eta) + 2 G_{\perp0}^2(2 + \eta(\eta(3 + 2\omega_r(3 + \omega_r)) - 5 - 4\omega_r)))\\
    + \eta^2(1 + \tau)(3(1 + \tau) - 2 G_{\perp 0})
\end{multline}

\bibliographystyle{jpp}

\bibliography{CaseVankampen}

\end{document}